\documentstyle[11pt]{article}
\def\RE{{\mbox{\rm I\kern-.21em R}}}
\def\ZZ{{\mbox{\small \sf Z\kern-.45em Z}}}

\begin{document}

{ \Large
\begin{center}

RESONANCE QUANTUM GATE\\
\vskip1cm \large S.\,Avdonin $^1$,\, N.\,Bagraev $^2$,
A.\,Mikhailova $^3$, B.\,Pavlov $^3$ \\ .\vskip10pt
\end{center}
}
\noindent $^1$ Department of Mathematical Science, University of Alaska, Fairbanks, 
AK 99775-6660, USA;\\ $^2$ A.F. Ioffe Physico-Technical Institute, St.
Petersburg, 194021, Russia.\\ $^3$ Laboratory of the Theory of
Complex Systems, V.A. Fock Institute of Physics,\\ St.Petersburg
State University,St.Petersburg, 198504, Russia;
\newtheorem{lem}{Lemma}[section]
\newtheorem{cor}{Corollary}[section]
\vskip0.5cm

\centerline{\large\bf Abstract} Simplest models of
two- and three-terminal Quantum Quantum Gates are suggested in
form of a quantum ring with few one-dimensional quantum wires
attached to it and several point-wise govering electrodes
inside the ring which are charged by a single hole. In
resonance case, when the Fermi level \footnote{See
\cite{Madelung}} in the wires coincides with the resonance energy
level on the ring the geometry of the device may be chosen such
that the quantum current through the switch from up-leading wire
to the outgoing wires may be controlled via re-directing of the
hole from one govering electrode to another one. The working
parameters of the gate are defined in dependence of the desired
working temperature, the Fermi level and the effective mass of
the electron in the wires. \vskip5pt

\section{Scattering problem and the switching effect}

Modern quantum electronic devices may be manufactured as
quantum networks of Aharonov-Bohm quantum rings and quantum wires
formed on the surface of a semiconductor. Elements of these
networks - rings and wires- may be obtained in self-assembled
quantum wells inside Silicon wafer via controlled diffusion of
borons with surface injection of vacancies,under proper
temperature, see for instance\cite{B1}, \cite{B2},\cite{B3}. The
width of the rings and wires $20$ A is small comparing with
De-Broghlie wavelength ( 80 - 150 A) on the Fermi level, hence
the network is quasi-one- dimensional.This permits to use the
Schr\"{o}dinger equation on the corresponding one-dimensional
graph as a convenient model when exploring the quantum
conductivity. The spectral theory of Schr\"{o}dinger equation on
the one-dimensional graph was developed in frames of the Theory
of Operator Extensions in \cite{GP}, \cite{Albeverio},
\cite{Extensions}, \cite{Novikov}, \cite{Schrader},\cite{Pasha}.
Using of this approach in mathematical design of nano-electronic
devices gives prediction of qualitative features of devices and
even preliminary estimation of their working parameters.
Interference of the wave functions served a base for design of the
first patented quantum switch, see \cite{ES}. Basic problems of
the mathematical design of quantum electronic devices were
formulated in terms of quantum scattering in the beginning of
nineties by \cite{A}. Design of most of modern resonance quantum
devices, beginning from classical Esaki diode up to the modern
devices, see, for instance ,\cite{Compano} were based on the {\it
resonance of energy levels} rather than on resonance properties of
the corresponding wave functions. At the same time modern
experimental technique already permits to observe resonance
effects caused by details of the shape of the resonance wave
functions, see \cite{B1}, \cite{B2}, \cite{B3}. In \cite{Ring} the
problem of {\it resonance manipulation of quantum current}
through the the quantum switch implemented as a quantum ring with
few incoming/outgoing quantum wires was considered in frames of
the relevant scattering problem. It was shown there, that the
transmission from one (incoming) wire to another (outgoing) wire
is defined not only by the position of the resonance eigenvalue
\footnote{Resonance eigenvalue is equal to the Fermi level in the
up-leading wires} on the ring, but also by the {\it shape of the
corresponding resonance eigenfunction }. In frames of the
EC-project "New technologies for narrow-gap semiconductors" (
ESPRIT-28890 NTCONGS,\,\,1998 - 1999) the problem on mathematical
design of a four-terminal Quantum Switch for triadic logic was
formulated by Professor G. Metakides and Doctor R.Compano from
the Industrial Department of the European Comission. Results of
the theoretical part of the project were published in papers
\cite{Ring}, \cite{Domain}, \cite{Solvay}. In \cite{P00},
\cite{PBar} \cite{MP00} a new design of Resonance Quantum
Switches (RQS) was suggested in form of a quantum domain or
quantum ring with a few quantum wires (terminals) attached to it.
The idea of the new design, as presented in \cite{P00} and
\cite{MP00} for the device designed in form of a quantum domain
(quantum well), is based on the phenomenon observed first, see
\cite{Opening}, in the scattering problem for acoustic scattering
on a resonator with a small opening: an additional term in the
scattering amplitude caused by the opening appeared to be
proportional to the value of the resonance eigenfunction at the
opening (for the Neumann boundary condition on the walls of the
resonator) or to the value of it's normal derivative (for the
Dirichlet boundary conditions), see \cite{MP01}.
\par
In \cite{MP00} the single act of computation - the switching
of the current - is considered
as a scattering process. In the corresponding mathematical
scattering problem for the RQS on the graph formed as a ring of
radius $R$ and a few (straight) quantum wires weakly connected to
it at the contact points $a_s$ via tunneling of electrons of
mass $m$ across the potential barriers width $l$, hight $H$, the
resonance eigenvalue $E_2 = E_f$ on the ring is embedded into
continuous spectrum of the Schr\"{o}dinger operator on the whole
graph. The width of the the part of the up-leading channel
connecting the wire and the well, and hence the power of the
potential barrier may be controlled by a special nano-electronic
construction -- the {\it split-gate}.On the other hand the power
of the potential barrier inside the split-gate may be manipulated
by the classical electric field applied to the brush of boron's
dipoles sitting on the shores of the channel. This construction
was suggested in experimental papers \cite{B1}, \cite{B2},
\cite{B3}, \cite{PBar}. The hight of the barrier, and hence the
power of it, is defined by the position of the lowest energy
level in the cross-section of the channel of variable width.
\par
If the potential barriers separating the wires from the
ring are
strong enough, then the connection between the wires and
the ring may be reduced to the boundary condition with
the small parameter
$$ \beta = \left(\cosh\frac{\sqrt{2m H}}{\hbar}l\right)^{-1} $$
at the contact points $a_s$. Here $H$ is the hight of the
barrier above the Fermi level and $l$ is the width of it. For
weak connection between the ring and the wires the transmission
coefficient from one wire to another in the resonance case appears
to be proportional to the product of the values of the normalized
resonance eigenfunctions $ \varphi(a_s)$ at the contact points
see \cite{Ring}. The resonance condition is fulfilled if the
re-normalized energy level $ E \to \lambda = k^2 = (E -
V_2)\frac{\hbar^2 R^2}{2m}$ ( proportional to the depth of the
quantum wires $E - V_2$ with respect to the resonance eigenvalue
$E$ ) is close to the re-normalized Fermi level $E_f \to \lambda_f =
(E_f - V_2)\frac{\hbar^2 R^2}{2m}$. If the connection between
the ring and the wires is weak,$\beta <<1$,then the
following approximate expression in scaled variables, see below
Section 3, is true for the transmission coefficient from the wire
attached to $a_s$ to the wire attached to $ a_t$ :
\[
S_{s,t} (\lambda) =
\frac{2k|\beta|^2}{k |\beta|^2
|\vec\varphi|^2 - i(\lambda_f - \lambda) } \varphi(a_s) \varphi
(a_t) + O (\beta^2), \,\, s\neq t.
\]
where $|\vec\varphi|^2$ is the length of the {\it
channel-vector} $\left(\varphi a_1,\,\varphi a_2,\,\dots \varphi
a_4 \right)$ the second term is uniformly small when $\beta\to 0$,
but the first one exhibits a nonuniform behaviour in dependence on
ratio $(\lambda_f - \lambda)/\beta^2$. The last formula being
applied formally to the case $\lambda = \lambda_f$
shows, that the transmission coefficient is approximately equal to
\[
S_{s,t} (\lambda) = \frac{2}{|\vec{\varphi}|^2}\varphi(a_s)
\varphi (a_t) + O(\beta^2).
\]
The transmission coefficient is not continuous with respect to the
re-normalized energy $\lambda$ uniformly in $\beta$. The
physically significant values of the transmission coefficient
for non-zero temperature $T$ may be obtained via averaging
over intervals $ |E - E_f|< \kappa T$ for {\it relatively small}
and {\it relatively large} temperature. In the first case we
still have:
\[
\overline{|S_{ij} (T)|^2} \approx
\frac{2\vert\varphi(a_s)\varphi(a_t)\vert^2} {|\vec{\varphi}|^4}
\]
but in the second case, we have :
\[
\overline{|S_{ij} (T)|^2} \approx
4\frac{\vert\varphi(a_s)\varphi(a_t)\vert^2}{|\vec{\varphi}|^4}
\frac{1}{1 + \frac{\kappa^2 T^2}{\lambda_0|\beta|^4
|\vec{\varphi}|^4 }}.
\]
Hence for small $\beta$ and non-zero temperatures the averaged
transmission coefficient is small, according to natural physical
expectations.
\par
Nevertheless the above formulae show that in certain range of
temperatures the transmission is proportional to the product
of values of the resonance eigenfunctions at the contact
points. Similar observation takes place for switches based on
the quantum well with Neumann boundary conditions ,see
\cite{MP00} and analog of it with normal derivatives of the
resonance eigenfunction remains true for the Dirichlet boundary
conditions, see \cite{MP01}.
\par
One may obviously construct the
{\it dyadic} RQS basing on this observation. But even {\it
triadic} (four-terminal) RQS may be constructed with minimal
alteration of the geometrical construction. For instance, on a
circular quantum well $\Omega_0 : |\vec x|\leq R$ the magnitude
of the constant electric field ${\cal E} \vec \nu,\,\, |\vec
\nu|= 1,$ and the shift potential $V_0$, see \cite{MP00}, may be
chosen such that the corresponding Schr\"{o}dinger equation
\begin{equation}
\label{Schred}
-\frac{\hbar^2}{2m}\frac{d^2 \psi}{d x^2} +
\left( {\cal E}e \langle \vec x,\, \vec \nu \rangle + V_0\right) \psi = E
\psi
\end{equation}
may have a resonance eigenfunction with an eigenvalue equal to
the Fermi level $E = E_{_f}$ in the wires such that it has only
one smooth line of zeroes which crosses the circle dividing it
in ratio $ 1:2 $. Then, attaching the wires $\Gamma$ to the domain
at the points $a_1,a_2,a_3,a_4$ characterized by the central
angles $ 0,\, \pm \pi/3,\, \pi $ we obtain the Resonance Quantum
Switch manipulated by rotating $\vec \nu \to \vec \nu'$ of the
constant electric field ${\cal E} \vec \nu$ in the plane
parallel to the plane of the device. In particular, if for some
direction $\vec\nu$ the line of zeroes arrives to the boundary
exactly at the contact points : $a_2,a_3$ (or $a_3,a_4$, or else
$a_4,a_2$ ) , then the corresponding wires $\Gamma_2,\, \Gamma_3$
( or respectively $\Gamma_3,\, \Gamma_4$ , or else $\Gamma_4,\,
\Gamma_2$) are blocked. With two outgoing wires blocked, one
up-leading wire $\Gamma_1$ and one outgoing wire (respectively
$\Gamma_4$, or $\Gamma_2$, or $\Gamma_3$) remain open. Hence the
electron current may go across the well from the up-leading wire
to the outgoing wire. The corresponding transmission and
reflection coefficients may be calculated in course of solution
of the corresponding scattering problem, \cite{MP00}.
\par
For fixed contact points the working regime of the
RQS is defined by the position of the working point $R,\,
{\cal E}, \, V_0 $ in the three-dimensional space of the
parameters. This position is uniquely defined, see Section $4$,
by the desired working temperature $T$ and by the Fermi level
$E_{_f}$ in the up-leading quantum wires. Note, that {\it the
position of the working point can't be defined experimentally }
just by naive scanning on one of parameters for other parameters
fixed at random, since the probability of proper choice of the
remaining parameters is zero (proportional to the zero-measure of
a point on a $2-d$ plane).
\par
In the contrary to the Quantum Switch, the Quantum Gate we
shall discuss is manipulated {\it not} by the classical electric
field, but by an {\it injection of a single quantum charged
particle} (a hole) into one of govering elecrtrodes situated
inside the ring. For the triadic (three-terminal) Resonance
Quantum Gate the problem of choosing of position of the govering
electrodes inside the ring is actually a sort of inverse problem
for the corresponding quantum network. This
problem may be formulated the following way :
\par
{\it Choose positions of the govering electrodes inside the
ring and the hight of them above the plane of the ring
such that \par 1. The resonance eigenfunction on the ring has
two zeroes which divide the ring in ratio $2\div1$. \par The
position of the zeroes may be controlled by the charging
of one of electrodes with a single hole such that the
zeroes of the resonance eigenfunction block two of outgoing
channels , leaving only one of outgoing channels open. }
\par
Solution of this inverse problem is non-unique. One of possible
solutions is suggested in the next section 3 of our paper.
\par
The plan of our paper is the following. In the second Section we
suggest a solution of a problem of the geometrical design of
a triadic Resonance Quantum Gate in a form of a circular quantum
well with three point-wise electrodes inside of it. In the
third section, following our previous results in \cite{RQSQG} we
define the small parameter in the boundary conditions.
In the last Section 4 we calculate the the working parameters of the
ring-based three-terminal Quantum Gate, manipulated by the
single-hole charging of electrodes situated inside the ring.
\par
We skip here discussion of two most important problems of the
mathematical design of the Resonance Quantum Switches and
Gates :
\par
The calculation of the Voltage-Current characteristics
and
\par
The estimation of affordable precision of the geometrical
details of it.
\par

\vskip5pt
\section{Resonance Quantum Gate}

The Resonance Quantum Switch manipulated by the macroscopic
electric field is actually a {\it classical device for
manipulating of quantum current}. It can't be used as a detail
of a quantum network since the macroscopic electric field can't
appear in the quantum network. In this section we consider a
completely quantum device manipulated by a single electron or
hole. Mathematical modeling of this device requires solution
of a two-electron problem on a network, similar to one solved
in the simplest case in \cite{MP95}. In this note we
consider a one-body version of the problem, assuming that a
single hole is sitting inside the circular ring at the hight
$h$ over some point $b_s$ on the continuation of one of the radii
corresponding to the contact point $a_s $ with outgoing wires at
the distance $|b_s| = b$ from the center of the ring. We may
have
three electrodes inside the ring, and following three possible
potentials which may be used for redirecting of the
electron current to three possible outgoing wires. The
Coulomb potential on the ring, produced by the single charge
sitting on one of electrodes, is, counting the charge of
electron,
\[
-\frac{e^2}{\sqrt{R^2 + h^2 + b^2 - 2 b R
\cos (\theta - \theta_s) }}.
\]
If the condition $\frac{2bR}{R^2 + h^2 + b^2} << 1$ is
fulfilled, then using the Taylor expansion we may find the
approximate expression for the potential energy of the
Schr\"{o}dinger equation on the ring in form :
\[
\label{Coulomb} V_s^C (x) \approx - \frac{e^2}{\sqrt{R^2 + h^2 +
b^2 }} -
\]
\begin{equation}
\frac{e^2 bR}{(R^2 + h^2 + b^2)^{3/2} } \cos(\theta - \theta_s) +
O(\frac{e^2 b^2 R^2}{(R^2 + h^2 + b^2)^{5/2}}).
\end{equation}
Neglecting the small addend we obtain the renormalized
harmonic potential in form
\[
V_s (x) = - Q \cos (\theta - \theta_s) - A ,
\]
and thus we arrive again to the Mathieu equation with the
potential $ {\cal E} e \langle \vec x,\vec \nu \rangle$ which
will be analysed in the last section . After introducing the
new variable
$z = 1/2 \,\,\,(\theta - \theta_s)$ the coefficients of the Mathiew
equation $$ y'' + (a - 2 q \,\,\, \cos(2 z)) y = 0 $$ are
calculated as
$$ a = \frac{8mR^2}{\hbar^2}\,
(E + \frac{e^2}{\sqrt{R^2 + h^2 + b^2 }}),$$
$$
q = - \frac{4 m e^2 bR^3}{\hbar^2 (R^2 + h^2 + b^2)^{3/2} }.
$$
Calculation of the working parameters will be accomplished in the
last part of the section 3 .
\par
The above one-body approximation may be used if the
life-time of the single hole on the electrode is greater
than the life-time of the resonance, but still small enough to
provide necessary speed of switching. If this condition is
not fulfilled, then the corresponding two-body
scattering problem should be analysed similarly to \cite{MP95}.

\section{ Boundary conditions at the contact points.}
\vskip5pt
In this Section, following \cite{RQSQG} we calculate
the parameter $\beta$ in the boundary conditions for the RQG based
on the quantum ring with few terminals.
We assume that the potential barrier
separating the ring from the quantum wire at the contact
point may be controlled by the split-gate mentioned above,
Section $1$, see also \cite{B2}, \cite{B3}.
\par
Consider a quantum gate constructed in form of a
circular ring of quasi-one-dimensional quantum wire
$\Gamma_0$ with a few straight radial up-leading wires
$\Gamma_s = \Gamma_{s1} \cup \Gamma_{s2}$ attached to it
orthogonally at the contact points $a_s,\, s = 1,2,\dots 4$.
The Schr\"{o}dinger equation on the ring $\Gamma_0$
is defined by some smooth potential $ q(x)+ V_0$, and the
Schr\"{o}dinger equations on the wires $\Gamma_s = \Gamma_{s1}
\cup \Gamma_{s2}$ have piecewise constant potentials
\[
V_s (x) = \left\{
\begin{array}{cc}
V_1 ,\,\, \mbox{if}& x \in \Gamma_1 :-l < x < 0,\\
V_2, \,\,\mbox{if}& x \in \Gamma_2 : 0<x<\infty.
\end{array}
\right.
\]
The point $x = -l $ on the wire $ \Gamma_s$ coincides with
the corresponding contact point $a_s$ on the ring $\Gamma_0$.
The hight $H = V_1 - E_f$ of the potential over the Fermi
level on the initial part of the quantum wire (within the
split-gate $x \in (-l, 0)$) is controlled by the electric field
orthogonal to the wire , which may change the width of the
the channel via turning the boron's dipoles sitting on the
shore of it \cite{B1},\cite{B2}. We assume, that the Fermi level
in the wires lies between $V_1, V_2$: $ H> 0> V_2 $ , the boundary
condition at the point of contact is chosen in Kirchhoff form
\footnote{In fact one may show that the boundary condition
connecting the solutions of the differential equations on the
wires and on the ring at the contact points depends on
local geometry of the joining. We consider the Kirchhoff
condition as a zero-order approximation for more realistic
boundary conditions.}:
\begin{equation}
\label{bc} [u'_0](a_s) + u'_{s}(a_s) = 0 ,
\end{equation}
and the solutions $u_{s} = \{ u_{s1},\, u_{s2}\} $ of the
Schr\"{o}dinger equations on $\Gamma_s = \{ \Gamma_{s1},\,
\Gamma_{s 2} \}$ are smooth functions on the joint interval
$(-l,0)\cup (0,\infty) = \Gamma_1 \cup \Gamma_2$ for which the
matching conditions are fulfilled
$$u_{s1}(-0) = u_{s2}(0),\,u'_{s1}(-0) = u'_{s2}(0).$$
\par
For ``relatively low" temperature $\kappa \Theta <
\frac{\delta_0}{2}$, see below , section 4, one may assume that
the dynamics of electrons is described as the restriction of the
evolution defined by the non-stationary Schr\"{o}dinger equation
onto the spectral interval length $\kappa T$ near the Fermi level
(that is near the corresponding resonance eigenvalue on the
ring). Practically we should calculate the scattering matrix on
the graph for values of energy inside this interval.Following
\cite{RQSQG} we shall use the Ansatz for the component of the
scattered wave on the ring in form of a linear combination
of the Green functions of a perturbed problem which takes
into account the presence of wires supplied with split-gates.
We assume that the potential of the split-gate is strong
enough $\frac{\sqrt{2m(V_1 - E_f)}}{\hbar} l = rl > 1$ and $e^{-
\frac{\sqrt{2m(V_1 - E_f)}}{\hbar} l}<< 1$. The presence of the
wires with split-gates may be modeled by additional singular
potential on the ring localized at the contact points. This
potential was calculated in \cite{RQSQG} from the Kirchhoff
conditions for the solutions $\psi, \psi_s$ of the Schr\"{o}dinger
equation on the ring and the Schr\"{o}dinger equation on the wires
at the contact points
\[
[\psi '](a_s) + \psi'_s \big|_{-l} = 0,\, \psi\big|_{a_s} =
\psi_s \big|_{-l}.
\]
If the potential barrier, defined by the split-gate in the
initial part of the wire is strong enough, $rl > 1$, then the
solution of the Schr\"{o}dinger equation with the constant
potential $V_1$ on the initial interval $(-l,0)$ of the wire
may be approximated by an exponential :
\[
\psi_s = C e^{-\frac{\sqrt{2m(V_1 - E_f)}}{\hbar} (x + l)}.
\]
Eliminating the Cauchy data of the decreasing exponential
solution on the wire we obtain from the above Kirchhoff condition
the the jumping boundary condition for the wave-function on the
ring at the contact points :
\[
[\psi '] - \frac{\sqrt{2 m (V_1 - E_f)}}{\hbar} \psi
\,\, \big|_{a_s} = 0.
\]
This boundary condition may be also presented in form of an
additional singular potential :
$$V(x) \longrightarrow V (x) +
\sum_{s = 1}^4 \delta (x - a_s) \frac{\hbar\sqrt{2m(V_1 -
E_f)}}{2m} := \tilde V (x).$$
Constructing of the Green function of the
``perturbed" Schr\"{o}dinger operator $$\tilde{L} =
-\frac{\hbar^2}{2m} \frac{d^2 \psi}{d x^2} + V(x) \psi + \sum_{s
=0}^{4} \delta (x - a_s) \frac{\hbar\sqrt{2m(V_1 -
E_f)}}{2m} \psi $$
on the ring with the new potential
$\tilde V$ serves as a convenient step to construct the scattered
waves on the whole graph.
\par
Note that the Kirchhoff boundary condition imposed at the contact point
$a$, which has the coordinate $x = -l$ on the corresponding
wire may be transformed to the point $x = 0$ of the wire with use
of the corresponding transfer-matrix for the Schr\"{o}dinger equation
on the wire $\Gamma_s$. It is convenient to use the scaled variables
on the ring $x\to \frac{x}{R}$ and on the wires $x_s \to \xi= \frac{x_s}{R}$
Denoting by $[\psi'_0],\, \psi$ the
symplectic variables of the component of the scattered wave
at the contact point $a_s/R$ on the scaled ring and by
$\tilde{\psi}_s,\,\,-\frac{d\tilde{\psi}_s}{d \xi}, $ the
symplectic variables of the component of the scattered wave on
the scaled wire at the end of the split-gate $\xi = 0$, we
obtain from the Kirchhoff condition the following boundary
condition:

\begin{equation}
\label{BoundCond}
\left(
\begin{array}{c}
[\frac{d\tilde{\psi}_0}{d\xi}] (a_s/R)\\
\tilde{\psi}_s (0)
\end{array}
\right)
==
\left(
\begin{array}{cc}
Rr \tanh rl & \frac{1}{\cosh rl}\\
\frac{1}{\cosh rl} & \frac{-1}{Rr} \tanh rl
\end{array}
\right)
\left(
\begin{array}{c}
\tilde{\psi}_0 (a_s /R)\\
- \frac{d\tilde{\psi}_s}{d\xi} (0)
\end{array}
\right).
\end{equation}
Denote by $\tilde{g} (\xi,\, \eta)$ the Green function of the
scaled perturbed Schr\"{o}dinger equation on the scaled ring. Then
the jump of the derivative of it is equal to $-1$ at all
points on the unit circle, except contact points,
where the jump is calculated as:
\[
[\tilde{g}']_{\xi = a_s/R} - rR \tilde{g}(a_s/R,\, a_s/R) = - 1.
\]
We choose an Ansatz for the component $\tilde{\psi}_0 (\xi)$
of the scattered wave on the ring in form
\[
\tilde{\psi}_0 (\xi) = \sum_{s = 1}^4 \tilde{g}(\xi,\, a_s/R)
u_0^s = \tilde{\bf g}(\xi) \vec{u}_0.
\]
Substituting now the corresponding component of the Ansatz
$\tilde{\bf g}({a_s}/R) \vec{u}_0 := \left\{\tilde{\bf g}
\vec{u}_0\right\}_s $ instead of $\tilde {\psi}_0 (a_s)$, and the
$s$-component of the standard Ansatz $ \left[ e^{ik\xi} + S(k)
e^{-ik\xi} \right] \vec{e}$ of the scattered wave for $\tilde
{\psi}_s$:
\[
\tilde {\psi}_s (0) = ( [I + S]\vec{e})_s := u_s (0),\,\,\,\,
\tilde {\psi}'_s (0) = ik( [I - S]\vec{e})_s := u'_s (0),
\]
into the above boundary condition we obtain the equation for
the Scattering matrix :
\begin{equation}
\label{BCcompl} \left(
\begin{array}{c}
-\vec{u}_0 \\
\vec{u} (0) -\frac{1}{Rr} \vec{u}' (0)
\end{array}
\right) = \left(
\begin{array}{cc}
Rr\frac{e^{- rl}}{\cosh rl} & \frac{1}{\cosh rl}\\
\frac{1}{\cosh rl} & - \frac{1}{Rr} \frac{e^{- rl}}{\cosh rl}
\end{array}
\right)
\left(
\begin{array}{c}
\tilde{\bf g} \vec{u}_0 \\ - ik [I - S]\vec{e}
\end{array}
\right),
\end{equation}
where $\tilde{\bf g}$ is a matrix combined of values
of the perturbed scaled Green function $\tilde{g}(a_s/R,\,a_t/R)$
at the contact points.
The scattering matrix may be found from these equations via
eliminating of the variables $\vec{u}_0$ on the ring.
\par
According to the above assumption the ratio $\displaystyle
1/\cosh rl := \beta$ may play the role of a small parameter in
the boundary condition . Then the diagonal elements of the matrix
in the right-hand side of the last equation have the second
exponential order, meanwhile the anti-diagonal elements are of the
first exponential order. Neglecting the elements of second order
we obtain the simplified version of the boundary
conditions
\begin{equation}
\displaystyle \label{SMBC}
\left(
\begin{array}{c}
-\vec{u}_0 \\\vec{u}(0)- \frac{1}{Rr} \frac{d \vec{u}}{d\xi}(0)
\end{array}
\right) = \left(
\begin{array}{cc}
0 & \frac{1}{\cosh rl}\\
\frac{1}{\cosh rl} &0
\end{array}
\right)
\left(
\begin{array}{c}
\tilde{\bf g} \vec{u}_0 \\ - \frac{d \vec{u}}{d\xi}(0)
\end{array}
\right),
\end{equation}
which coincides with the phenomenological boundary condition
we used in the section 1 for calculating of the scattering
matrix.
\par
Taking into account that $\vec{u}(0) = \frac{ik(I - S)}{\cosh rl}
\vec{e} $ we may solve the equation (\ref{SMBC}) with respect to
the $S \vec{e}$ for any $4$-vector $\vec{e}$ and then obtain an
expression for the scattering matrix in form
\begin{equation}
\label{SM} \displaystyle
S(k) = \frac {\frac{\tilde{\bf g}(\lambda)}{\cosh^2 rl}
+ \frac{1}{Rr} - \frac{1}{ik}}
{\frac{\tilde{\bf g}(\lambda)}{\cosh^2 rl} + \frac{1}{Rr} + \frac{1}{ik}}.
\end{equation}
Here $\tilde{\bf g}(\lambda)$ is a matrix combined of values
at the contact points of the Green- function of the scaled
perturbed equation on the ring: $\left\{\tilde{\bf g} (\lambda)
\right\}_{st} = \tilde{g}(a_s/R,\, a_t/R, \, \lambda),\, \lambda =
k^2 $.
Now, similarly to analysis done in \cite{Ring} we
may separate the leading terms in the numerator and denominator of
the matrix function $\tilde{\bf g} (\lambda)$ near the resonance
eigenvalue $\mu_2$ of the perturbed operator :
\[
\tilde{\bf g} = \frac{|\vec{\varphi}_2|^2 P_2}{\mu_2 - \lambda} +
K.
\]
Here $P_2$ is the orthogonal projection in $4$-dimensional
euclidean space onto the vector $\vec{\varphi}_2$ formed of the
values of the resonance eigenfunction ${\varphi}_2$ at the contact
points; the non-singular addend $K$ may be estimated
similarly to \cite{Ring}. If the condition of domination of the
non-singular term $K$ by the group of the
leading terms,is fulfilled, in a small real neighborhood of the resonance
eigenvalue $\mu_2$, then one may pertain the leading terms only
when calculating an approximate expression for the scattering
matrix in this neighborhood :
\begin{equation}
\label{simple} \displaystyle S_{approx}(k) = \frac{
\frac{|\vec{\varphi}_2|^2}{\mu_2 - \lambda} P_2 + \cosh^2 rl
(\frac{1}{R r} - \frac{1}{ik})} {\frac{|\vec{\varphi}_2|^2}{\mu_2
- \lambda} P_2 + \cosh^2 rl (\frac{1}{R r} + \frac{1}{ik})}.
\end{equation}
This gives actually a convenient two-poles approximation for
the scattering matrix and estimation of life time of resonances-
the speed of the decay of the resonance
terms in solution of the non-stationary Schr\"{o}dinger
equation.It suffice to calculate zeroes of the leading term
in the numerator assuming that $\cosh rl >> 1$. Using the
notation $\alpha = \pm \sqrt{\mu_2}$,
we obtain two zeroes in lower half-plane :
\begin{equation}
\label{appzero}
k \approx \alpha + \frac{irR }{ i\alpha - rR}\,\,\,
\frac{|\vec{\varphi}_2|^2}{2 \cosh^2 rl}.
\end{equation}
The two-poles approximation follows from it , and the imaginary
parts of zeroes define the inverse life time $\tilde{\gamma}$ of
the resonances. The scaled time $\tau $ corresponding to the
scaled equation and the real time $t$ are connected by the
formula $k^2 \tau = (E - V_2)t$, or $t = \frac{2mR^2}{\hbar^2}\tau
$. The exponential decay of the resonance states of both scaled
and the non-scaled equations is defined by the behaviour of he
exponential factor $e^{i k^2 \tau} = e^{i \Re k^2 \tau} e^{-\Im
k^2 \tau}$. The decreasing exponential factor may be rewritten
with respect to real time as $e^{\Im k^2 \frac{\hbar^2}{m R^2}
t}$. Hence the role of the real inverse life time is played by
$\Im k^2 \frac{\hbar^2}{m R^2} $ and may be calculated
approximately for $\cosh rl >>1$ as
\begin{equation}
\label{lifetime}
\frac{\alpha r^2 \hbar^2 |\vec{\varphi}_2|^2}{2 m
(\alpha^2 + r^2 R^2) \cosh^2 rl}.
\end{equation}
For intermediate values of $\cosh^2 rl >> \frac{1}{ \sqrt{1 +
\frac{r^2 R^2 }{\mu_2}}}$ one may simplify the expressions in
both terms of the previous formula for the scattering matrix
neglecting $e^{-2rl}$ compared with $ \sqrt{1 + \frac{r^2 R^2
}{\mu_2}}$. Then we obtain another convenient approximate
expression for the {\it approximate scattering matrix} near the
resonance eigenvalue $\mu_2$:
\[
S_{approx} (k) = \frac{3ik - Rr}{3ik + Rr}\,\,\,\,\frac{2\cosh^2
rl \frac{Rr - ik}{Rr - 3ik} + Rr \tilde {g}}{2\cosh^2 rl \frac{Rr
+ ik}{Rr + 3ik} + Rr \tilde {g}}.
\]

We shall discuss now a version of RQG
based on a circular {\it quantum ring} $\Gamma_{_0}$ of radius
$R$ with three {\it outgoing} straight radial quantum wires
$\Gamma_{_s},\, s= 1,2,3 $ attached to it at the points
$\varphi = \pm \pi/3,\, \pi$ via tunneling across the potential
barriers controlled by the split-gates. We assume, that the
up-leading quantum wire is supplied with so high potential
barrier that the jump of the derivative of the
wave-function on the ring at this point may be neglected
when calculating the eigenvalues and eigenfunctions of the
perturbed operator $\tilde{L}$. Still we pertain the jumps at
the contact points of the outgoing channels $\varphi = \pm
\pi/3,\, \pi$, characterized by the potential barriers width
$l$ and hight $H = V_1 - E_f$ over the Fermi level $E_{_{f}}$ in
the radial quantum wires $\Gamma_{_s}$. In this Section we
assume that RQS is manipulated by the constant macroscopic
electric field ${\cal E} \vec \nu $ which generates the
potential ${\cal E} e R \langle \vec \nu,\,\vec x \rangle + V_0
$ in the Schr\"{dinger} equation on the ring $\vec x = R \vec
\xi , |\vec \xi| = 1 $. It is equivalent to the leading term of
the electrostatic potential generated by the single hole
sitting on the selected electrode. We
assume as before that the influence of the field on the quantum
wires is eliminated by some additional construction, so that the
potential on the wires produced by the above field is equal to
zero. It means that the Schr\"{o}dinger equation on the network
combined of the up-leading wire, the ring $\Gamma_{_0} $ and
outgoing wires $ \Gamma_{_1},\, \Gamma_{_2},\,\Gamma_{_3}\,$ may
be written as a system of Mathieu equation on the ring
$\Gamma_0$
\[
- \frac{\hbar^2}{2m}\frac{d^2 u}{d x^2} +
\left({\cal E} e \langle \nu,\,x \rangle +
V_0 \right) u = E u,\,\,\, u = u_0,
\]
and the Schr\"{o}dinger equation with the step-wise potential
\[
V_{_s} (x) =
\left\{
\begin{array}{cc}
H + E_{_f}, & \mbox{\,\, if } - l < x < 0,\\ E_{_f} + V ,
&\mbox{\,\, if \,\,} 0 \leq x < \infty,\,\, V <0
\end{array}
\right. :
\]
\[
- \frac{\hbar^2}{2m}\frac{d^2 u}{d x^2} +
V_{_s} (x) u = E u,\,\,\, u = u_s,
\]
on the outgoing wires $\Gamma_{_s}, \, s=1,2,3$ and on the
incoming wire $\Gamma_{_4}$. We assume also, that the incoming
wire $\Gamma_{_4}$ is attached to the quantum ring at some point
$a$ which is different from the above points $a_s$. The connection
of the outgoing wires with the quantum ring is characterized by
the ``small" parameter $\beta =(\cosh \frac{\sqrt{2m (V_1 -
E_f)}}{\hbar} l )^{-1} <<1.$
\par

An important engineering problem is actually {\it the proper
choice of the electric field } ${\cal E}$ such that the
corresponding differential operator on the ring has an
eigenfunction with special distribution of zeroes: the zeroes
of the eigenfunction corresponding to the second smallest
eigenvalue should divide the ring in ratio $1\div2$. We
assume that the potential barrier at the contact point
$a_{_4}$ with the incoming wire is so high that we may
neglect the jump of the derivative of the perturbed operator
$\tilde{L}$ eigenfunction at this point. Then the whole potential of
$\tilde{L}$
on the ring is combined of the smooth potential defined by
the macroscopic field ${\cal E}$ and an
additional singular potential appearing from the Kirchhoff's
conditions of smooth matching of the solution $\psi$ at
the contact points $a_s,\, s= 1,2,3$ on the ring
with proper solutions of the
equations on the wires when the energy is fixed on the
Fermi level $ E = E_f$ :
\[
[\psi'_0] - \frac{\sqrt{2m (V_1 - E_f)}}{\hbar} \psi\,\,
\bigg|_{a_{s}} = 0,
\]
\[
-\frac{\hbar^2}{2m} \frac{d^2 \psi}{d x^2} +
{\cal E} e \langle \vec x,\vec \nu \rangle +
\sum_{s = 1}^N \delta (x - a_s) \frac{\hbar\sqrt{2m(V_1 -
E_f)}}{2m}\psi + \left( V_0 - V_2 \right) \psi= E \psi.
\]
We select the field ${\cal E}$ , and hence the geometry of
the device such that the resonance eigenfunction for $E = E_f$
would have, for certain direction of the unit vector $\vec \nu
$, two zeroes on the ring sitting at the points $\varphi =
\pm\pi/3$ . When
using the standard form of the scaled Mathieu equation
with properly re-normalized coefficients
$q = \frac{4m {\cal E}e R^3}{\hbar^2},\,
a = frac{8m R^2 (E-V_0 + V_2)}{\hbar^2} $,
\begin{equation}
\label{standard}
y'' + (a - 2 q \,\,\, \cos(2 z)) y = 0,
\end{equation}
we should pass from the angular (scaled) variable
$\xi =x / R $ to the new variable $z = \frac{x}{2R} $ which is
changing on the interval $-\pi/2,\, \pi/2$. We have found that
if the vector $\vec \nu$ is directed toward $ z = 0$, and the
solution $\psi$, we are looking for, is an even (cosine-type )
solution of the Mathieu equation on the scaled ring $-\pi/2 < z
< \pi/2$ with a positive value at the point $ z = 0$, and zeroes
at $z = \pm \pi/6 $, then it is smooth at the contact points
with $z = = \pm \pi/6$ and has a jump of the derivative
$[\tilde {\psi}']_{\pi} = \frac{R\sqrt{2m (V_1 - E_f)}}{\hbar}
\tilde{\psi} (\pi) $ at the point $\xi = \pm\pi/2$. Hence, $y =
\tilde{\psi}(2z) $ satisfies the Mathieu equation on the interval
$(0,\pi /2)$ and the boundary conditions
\begin{equation}
\label{resbc}
y' (0) =
0, $$ $$
\frac{d y}{dz} (\pi/2) + \frac{R\sqrt{2m (V_1 -
E_f)}}{\hbar} y (\pi/2) = 0.
\end{equation}
The dimensionless Mathieu equation in standard form
(\ref{standard}) with properly scaled variable $z$ ,
$-\pi/2 < z <\pi/2 $, was analyzed with Mathematica
in dependence of the re-normalized electric field
and the parameter $ \displaystyle \gamma = \frac{R\sqrt{2m (V_1
- E_f)}}{\hbar}$ in the boundary condition
$[\frac{d\tilde{\psi}}{d\xi}] - \gamma \tilde{\psi} = 0 $ at the contact
points. It was found that for the following values of the
parameters $q,\, \gamma $ the resonance eigenfunction with two
zeroes at $ z = \pm \pi/6 $ exists, for instance :
\[
\gamma = 10,\,\, q = -1.98,\,\, a = 5.24.
\]
For the parameter $q$ selected as shown above, there exist an
eigenfunction $u$ of the Mathieu equation perturbed by the
$\delta$-potentials attached to the points $a_s$ with weight
$\gamma$ such that the zeroes of $u$ divide the unit circle in
ratio $1 \div 2$. These eigenfunctions may play a role of the
resonance eigenfunctions for the corresponding triadic Resonance
Quantum Gate. Being normalized by the condition $\varphi(0) = 1 $
this function has square $L_2$-norm
$3.5234 $. The spacing between the resonance
eigenvalue $\mu_2 = \frac{a}{4} = 1.30$
on the unit ring and the nearest eigenvalue
$\mu' =\frac{a'}{4}= 1.49 $ (from
the odd series of the eigenfunctions) is estimated as $0.19$.
Now the working temperature of the switch may be estimated
as in the next Section 4: $\kappa T \leq \frac{0.19\,\,\hbar^2}{2m
R^2}$. For quantum rings with radius $10$ nm the switching
time estimated from the life time of the corresponding
resonance may be circa $10^{-17}$ sec.
\vskip1cm
\section{High-temperature triadic RQG }
\vskip5pt
Consider a RQG constructed in form of a quantum domain -
a circular quantum well - with four terminals - quantum wires -
attached to it at the contact points $a_1,\, a_2,\, a_3,\, a_4 ,$
selected as suggested above. To choose the working point of
the switch in dependence of desired temperature we consider first the
{\it dimensionless} Schr\"{o}dinger equation
\begin{equation}
\label{dimless}
-\bigtriangleup u + \epsilon \langle \vec \xi ,\, \vec \nu \rangle u =
\lambda u
\end{equation}
in the unit disc $|\vec \xi| < 1$ with Neumann or Dirichlet
boundary conditions at the boundary. The dimensionless
Schr\"{o}dinger equation may be obtained from the
original equation by scaling $\vec x = R \vec \xi$ :
\begin{equation}
\label{connect}
-\bigtriangleup_{\xi} u +
\frac{2 m e {\cal E}R^3}{\hbar^2}\langle \vec \xi ,\, \vec \nu\rangle u =
\frac{2m R^2 (E - V_0)}{\hbar^2}u.
\end{equation}
Here ${\cal E}$ is the magnitude of the selected electric
field and the unit vector $\vec \nu$ defines it's direction,
$e$ is the absolute value of the electric charge of the
electron and $R$ is the radius of the circular well. Selecting $
\displaystyle \epsilon = \frac{2me {\cal E}R^3}{\hbar^2} =3.558$
for Neumann boundary conditions one may see, \cite{MP00}, that
the eigenfunction corresponding to the second lowest eigenvalue
$\mu_2 = 3.79$ of the dimensionless equation (\ref{dimless}) has
only one smooth zero line in the unit disc which crosses the
unit circle at the points situated on the ends of radii
forming the angles $\pm \frac{\pi}{3}$ with the electric field
$\epsilon \vec \nu$. The minimal distance $\delta_0$ of $\mu_2$
to the nearest eigenvalues (the spacing of eigenvalues at
$\mu_2$), depending on boundary condition on the border of
the well, may be between $2$ and $10 $.
For Dirichlet or Neumann boundary conditions the
eigenfunctions of the spectral problem for the above
Schr\"{o}dinger equation (\ref{dimless}) are even or odd
with respect to reflection in the normal plane containing
the electric field $\epsilon \vec{\nu}$. In particular for
the Neumann boundary conditions the nearest eigenvalues
corresponding to {\it even } eigenfunctions are equal $ \mu_1 = -
0.79$ and $\mu_3 = 9.39$, that it the spacing between $\mu_2$
and other eigenvalues of the {\it even} series may be
estimated as $\delta_0 := $ min$\{|\mu_2 - \mu_1|,|\mu_2 -
\mu_3|\} \approx 4$. Generally for the circular domain the
spacing between the second lowest eigenvalue $\mu_2$ and other
eigenvalues ( of both even and odd series ) may be estimated
from below as $\delta_0 \geq 2$. The working regime of the
switch will be stable if the bound states corresponding to
the neighboring eigenvalues will not be excited at the
temperature $T$ :
\begin{equation}
\label{Temper}
\displaystyle \kappa T \,\,\, \frac{2m R^2}{\hbar^2} \leq
\frac{\delta_0}{2}.
\end{equation}
This condition may be formulated in terms of the {\it
scaled temperature } $\displaystyle \Theta = \frac{2mR^2
T}{\hbar^2}$ as
\begin{equation}
\label{reduced} \kappa\Theta < \frac{\delta_0}{2}.
\end{equation}
The temperature which fulfils the above condition we call
{\it low} temperature for the given device. If the radius $R$
of the corresponding quantum well is small enough , then it
may work at the (absolutely) high temperature, which correspond
to the {\it low} scaled temperature. It may take place if
the radius of the well is sufficiently small. Importance
of developing technologies of producing devices of small size
with rather high potential barriers is systematically underlined
when discussing the prospects of nano-electronics, see for
instance \cite{Compano}.
\par
We assume that the effective depth $V_f$ of the bottom
value $V_2$ of the potential on the wires from the Fermi-level
$E_f$ in the wires is positive $V_f = E_f - V_2 > 0 $, and the
De-Broghlie wavelength on Fermi level is defined as $$
\displaystyle \Lambda_f =\frac{h}{\sqrt{2m V_f}}.$$ Then we
obtain the estimate of the radius $R$ of the domain from
(\ref{Temper}) as:
\begin{equation}
\label{R} \frac{R}{\Lambda_f}\leq \sqrt{\frac{V_f}{\kappa T}}
\sqrt{\frac{\delta_0}{8 \pi^2}}.
\end{equation}
For fixed radius $R$, the shift potential $V_0$ may be defined
from the condition
\[
\frac{2m R^2 [E_f - V_0]}{\hbar^2} = \mu_2.
\]
For instance, if we choose the radius $R$ of the domain
as $\displaystyle R^2 = \frac{\delta_0 \hbar^2}{4 m \kappa T}$,
we obtain :
\[
V_0 = E_f - \frac{\hbar^2 \mu_2 }{2 m R^2} = E_f - 2\kappa
T\frac{\mu_2}{\delta_0}.
\]
Finally, the electric field ${\cal E}$ may be found from
the condition
\[
\epsilon = 3.8 = e {\cal E}\frac{2m R^3}{\hbar^2},
\]
where $e$ is the absolute value of the electron charge. Hence
for the value of $R$ selected above we have :
\[
e{\cal E} R = \epsilon \frac{\hbar^2 }{2 m R^2} = \frac{2
\epsilon \kappa T}{\delta_0}.
\]
Hence the switch may work even at room temperature if the
radius $R$ of the quantum well is small enough and the
geometrical details are exact.
\par
Similar calculations may be done for the circular quantum
well with Dirichlet boundary conditions. It appeared that for
the dimensionless equation with the potential factor $\epsilon
= 18.86$ the eigenfunction with a single zero-line dividing the
unit circle into ratio $1:2$ corresponds to the second smallest
eigenvalue $\mu_2 = 14.62 $. The lowest eigenvalue which
corresponds to the {\it even} eigenfunction is $\mu_1 = 2.09$ and
the spacing between $\mu_2$ and other eigenvalues of both
even and odd series is estimated as before, $\delta_0 \geq
2 $. This gives proper base for calculation of the radius of the
quantum well,the intensity of the electric field and the shift
potential subject to given temperature and the Fermi level.

\end{document}